\documentclass[                  
 	,draft
               ]{aipproc}
               
\layoutstyle{6x9}

\usepackage{graphicx}
\usepackage{latexsym}
\usepackage{amsfonts}
\usepackage{color}
\usepackage{colortbl}
\usepackage{amssymb,amsmath,enumerate,undertilde}
\usepackage{amsthm} % AMS package for THEOREM environment
\usepackage{bm}
\usepackage{indentfirst}
\usepackage{mathrsfs}
\usepackage[T1]{fontenc}
\usepackage{psfrag}

\newcommand{\Real}{\mathbb{R}}

\begin{document}
%%%%%%%%%%%%%%%%%%%%%%%%%%%%%%%%

\title{The propagation of particles and fields in wormhole geometries}

\classification{04.20.-q,04.25.-g,04.40.-b}
% 04.20.-q: Classical GR
% 04.25.-g: Approximation methods; equations of motion
% 04.25.Dm: Numerical relativity
% 04.40.-b: Self-gravitating systems, continuous media and 
%           classical fields in curved spacetime
% 04.70.-s: Physics of black holes
% 97.60.Lf: Astronomy: Late stage of star evolution: black holes
% Complete list see \texttt{http://www.aip..org/pacs/index.html}
           
\keywords{general relativity, wormholes, stability analysis}

\author{Olivier Sarbach}{
address={Instituto de F\'{\i}sica y Matem\'aticas,
Universidad Michoacana de San Nicol\'as de Hidalgo\\
Edificio C-3, Ciudad Universitaria, 58040 Morelia, Michoac\'an, M\'exico.}
}

\author{Thomas Zannias}{
address={Instituto de F\'{\i}sica y Matem\'aticas,
Universidad Michoacana de San Nicol\'as de Hidalgo\\
Edificio C-3, Ciudad Universitaria, 58040 Morelia, Michoac\'an, M\'exico.}
}

%%%%%%%%%%%%%%%%%%%%%%%%%%%%%%%%
\begin{abstract}
We discuss several properties of static, spherically symmetric wormholes with particular emphasis on the behavior of causal geodesics and the propagation of linear fields. We show there always exist null geodesics which are trapped in a region close to the throat. Depending upon the detailed structure of the wormhole geometry, these trapped geodesics can be stable, unlike the case of the Schwarzschild black hole. We also show that test scalar fields propagating on such wormholes are stable. However, when a mixture of ghost and Klein-Gordon scalar fields is used as a source of the Einstein equations we prove that the resulting static, spherically symmetric wormhole configurations are linearly unstable.
\end{abstract}

\maketitle

%%%%%%%%%%%%%%%%%%%%%%%%%%%%%%%%%%%%%%%%%%%%%
\section{Introduction}
\label{Sect:Intro}
%%%%%%%%%%%%%%%%%%%%%%%%%%%%%%%%%%%%%%%%%%%%%

Traversable wormholes has been the subject of numerous investigations over the last few decades. Whether they are considered as a spacetime consisting of two asymptotically flat ends connected by a throat or as a handle connecting two distinct regions of the same universe, wormholes exhibit  properties that fascinate the public and scientists alike: transport of matter from one asymptotically flat end to the other and interstellar travel, the existence of close timelike curves and the possibility of backwards time travel, etc., see for example~\cite{mMkT88,mMkTyU88,Visser-Book}.

Although there exist many spacetimes describing traversable wormholes, in the context of General Relativity, a severe restriction comes from the requirement that these spacetimes satisfy Einstein's field equations. This leads to the necessity of exotic matter, ie. a configuration whose stress energy tensor violates the null energy condition~\cite{mMkT88,jFkSdW93}, and presently this type of matter is a hypothetical rather than a configuration supported by experimental evidences. On the other hand, the observed late time accelerated expansion of the universe seems to demand some form of dark energy which could, in principle, be modeled by exotic matter, although it is not clear whether or not such expansion really requires matter violating the null energy condition. Therefore, so far neither cosmological observations nor terrestrial physics  provide direct support for the existence of exotic matter. 

However, it has been known long ago~\cite{hEvGaJ65} that within the framework of Quantum Field Theory the renormalized stress energy tensor allows the violations of all so far known energy conditions and this leaves open the possibility that exotic matter may be provided by quantum fields. These expectations have faded away after the development of the quantum inequalities~\cite{lF78, lF91, lFtR96} that put strong constraints on the size of the region where violations of the  energy conditions occur. For further discussions on this important issue and the relevance of this type of analysis for wormholes, see for example~\cite{uY05, eFrW96, cFtR05}. Often in the literature the existence of exotic matter is circumvented by employing generalized theories of relativistic gravity such as theories arising as the low energy limit of string theories. Examples are dilatonic Gauss-Bonnet theory or other higher curvature theories like $f(R)$ theories. In these generalized theories wormholes have been constructed without any need for exotic matter (see for example~\cite{pKbKjK11, pKbKjK12, nMfL11}).

However, irrespectively of the issue regarding the existence and nature of the exotic matter, an important open problem concerns the stability of wormholes. Does there exist a matter model in the context of General Relativity which admits stationary wormhole solutions which are stable with respect to small perturbations of the metric and matter fields? So far stability analyses have been mainly restricted to spherically symmetric wormholes. For the case of a massless ghost scalar field, it has been shown in~\cite{jGfGoS09a} that all the static and spherically symmetric wormholes are linearly unstable, and a numerical evolution of the spherical nonlinear field equations reveals that these wormholes either expand or collapse to form a black hole~\cite{hSsH02,  jGfGoS09b, aDnKdNiN08}. Charged generalizations of these wormholes were also considered in~\cite{jGfGoS09c} and shown to be linearly unstable. Further analysis includes a model involving exotic dust in combination with a radial magnetic field~\cite{aSiNnK08, dNaDiNaS09} where linearly stable wormholes are obtained. However, as was later shown in \cite{oStZ10}, these configurations are unstable due to the formation of shell crossing singularities. Very recently, a generalization of the previous model to an exotic perfect fluid with pressure has been considered and claimed to yield wormholes which are stable with respect to radial perturbations~\cite{iNaS12}. It would be very useful to have a more general understanding on the stability issue of wormholes.

The present work discusses a few properties of wormhole spacetimes which we believe may be of relevance to the stability problem. We start in the next section with an analysis of the asymptotic behavior of the causal geodesics on a given, static and spherically symmetric wormhole background. We show that the presence of the throat leads to the existence of trapped timelike and null geodesics orbiting around the throat, and depending upon the nature of the effective potential these geodesics can be stable in the sense that a small perturbation of the initial data belonging to these curves leads to the same trapped behavior. Next, we study the time evolution of test scalar fields on the wormhole background and show that they remain bounded in time. Finally, we consider a matter model consisting of a mixture of ghost and Klein-Gordon scalar fields, and ask whether or not the inclusion of the Klein-Gordon fields could act as a stabilizer agent. However, we find that in this model all the static, spherically symmetric wormholes are linearly unstable.

%%%%%%%%%%%%%%%%%%%%%%%%%%%%%%%%%%%%%%%%%%%%%
\section{Geodesic motion on a wormhole background}
\label{Sect:Geodesics}
%%%%%%%%%%%%%%%%%%%%%%%%%%%%%%%%%%%%%%%%%%%%%

In this section we analyze geodesic motion on a given wormhole spacetime of the form
\begin{equation}
M = \Real^2 \times S^2,\qquad
{\bf g} = -dt^2 + dx^2 + r(x)^2\left( d\vartheta^2 + \sin^2\vartheta\; d\varphi^2 \right),
\label{Eq:WHMetric}
\end{equation}
where here $(t,x)$ are Cartesian coordinates on $\Real^2$, $(\vartheta,\varphi)$ are the standard polar coordinates on the unit $2$-sphere $S^2$, and $r(x)$ is a strictly positive smooth function of $x$ satisfying $\lim\limits_{x\pm \infty} r(x)/|x| = 1$. The function $r(x)$ has a global minimum at $x_{th}$ which describes the wormhole throat whose area is $4\pi r(x_{th})^2$ and which connects the two asymptotically flat ends at $x\to +\infty$ and $x\to -\infty$. A particular example is $r(x) = \sqrt{x^2 + b^2}$, with $b > 0$, corresponding to the Bronnikov-Ellis solution~\cite{hE73,kB73} whose properties in the context of wormholes were discussed in \cite{mMkT88}.

The causal geodesics $\gamma(\lambda)$ on the spacetime~(\ref{Eq:WHMetric}) are described by the Lagrangian
\begin{displaymath}
L = {\bf g}(\dot{\gamma},\dot{\gamma}) = -\dot{t}^2 + \dot{x}^2 + r(x)^2\dot{\varphi}^2,
\end{displaymath}
with a dot denoting differentiation with respect to an affine parameter $\lambda$. Here, we assume without loss of generality that the motion takes place in the equatorial plane $\vartheta = \pi/2$. Since $L$ does not depend explicitly on $t$ nor on $\varphi$, the quantities $E := \dot{t}$ and $\ell := r(x)^2\dot{\varphi}$ are constant along the trajectories. This leads to the effective mechanical problem
\begin{equation}
\dot{x}^2 + V(x) = E^2 - k^2,\qquad
V(x) = \frac{\ell^2}{r(x)^2},
\label{Eq:Geodesic}
\end{equation}
with $k^2 = 0$ for null and $k^2 > 0$ for timelike geodesics. A rescaling $\lambda\mapsto A\lambda$ of the affine parameter by a positive constant $A$ implies the transformations $E\mapsto E/A$, $\ell\mapsto \ell/A$ and $k\mapsto k/A$ of the conserved quantities.

\subsection{Null geodesics}

For null geodesics, the rescaling freedom mentioned above allows us to choose $E = 1$, and Eq.~(\ref{Eq:Geodesic}) can be rewritten as
\begin{equation}
\dot{x}^2 + \frac{\ell^2}{r(x)^2} = 1.
\label{Eq:NullGeodesics}
\end{equation}
Therefore, we are looking for the trajectories with energy one in the effective potential $\ell^2/r(x)^2$ which scales like the square magnitude of the angular momentum.
This potential is positive, has a global maximum at the throat, $x=x_{th}$, and decays to zero as $\ell^2/x^2$ for $x\to \pm\infty$.

In order to study the asymptotic behavior of the null geodesics, we consider an event $p\in M$ and a future-directed null vector ${\bf l}\in T_p M$, and examine the asymptotic limit for $\lambda\to\infty$ of the null geodesic through $p$ with tangent ${\bf l}$ at $p$. We may parametrize ${\bf l}$ in terms of the angle $\alpha$ according to
\begin{displaymath}
{\bf l} =  \frac{\partial}{\partial t} + \cos\alpha\frac{\partial}{\partial x}
 + \frac{\sin\alpha}{r(x_0)}\frac{\partial}{\partial \varphi} \; ,
\end{displaymath}
where $x_0$ is the $x$-coordinate of $p$ and the normalization has been chosen such that $\dot{t} = E = 1$. The condition $\dot{\varphi} = \ell/r(x_0)^2$ implies that 
\begin{equation}
\ell = r(x_0)\sin\alpha.
\label{Eq:ell}
\end{equation}

\begin{figure}[ht]
\centerline{\resizebox{10cm}{!}{\includegraphics{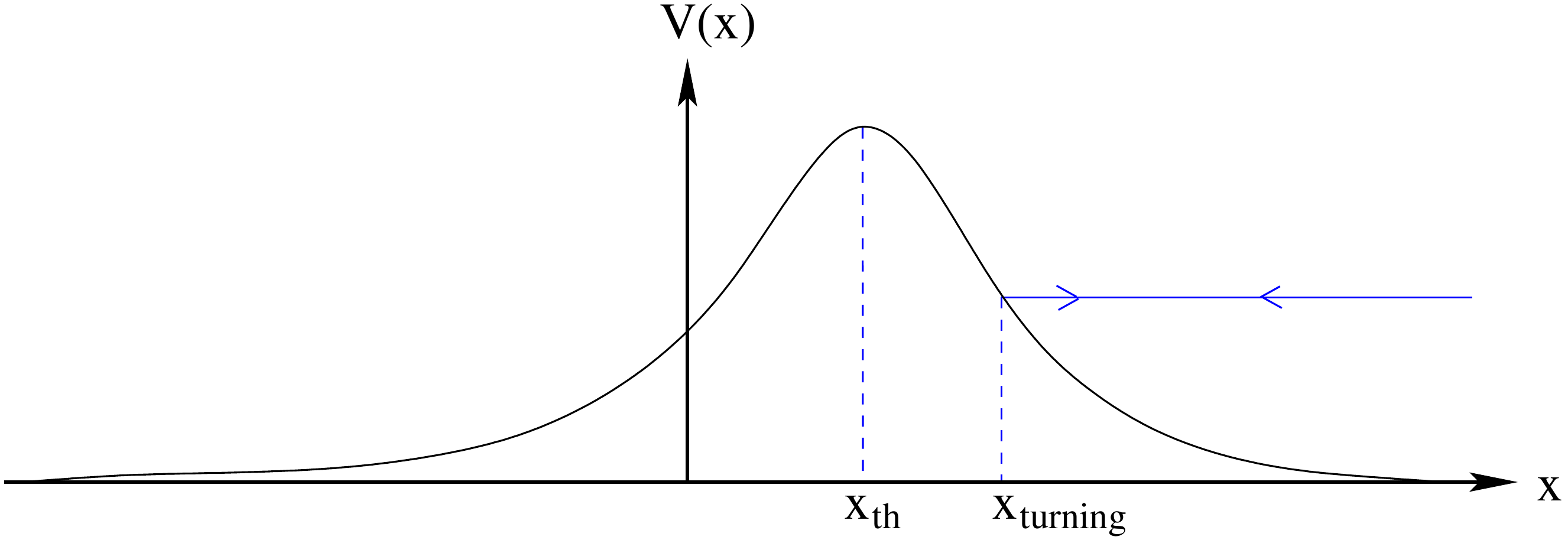}}}
\vspace{0.1cm}
\caption{The structure of the effective potential assuming that $r(x)$ is strictly convex. The unique maximum corresponds to unstable circular orbits.}
\label{Fig:Potential}
\end{figure}

Before we proceed with the analysis, let us assume first that the areal function $r(x)$ is strictly convex, as is the case for the Bronnikov-Ellis wormhole $r(x) = \sqrt{x^2 + b^2}$ mentioned above. This implies that the effective potential $V(x) = \ell^2/r(x)^2$ has a unique maximum at the throat $x = x_{th}$, see Fig.~\ref{Fig:Potential}.
With this assumption, we arrive at the following conclusions:
\begin{enumerate}
\item[(i)] The radial null geodesics correspond to the choice $\alpha=0$ or $\alpha = \pi$ for photons moving towards $x\to +\infty$ and $x\to -\infty$, respectively. In this case, the angular momentum $\ell$ is zero, and  thus the photons reach their respective asymptotic regions.
\item[(ii)] This behavior persists for angles $\alpha$ close to $0$ or $\pi$, as long as the maximum of the effective potential, $\ell^2/r(x_{th})^2$, lies below one. In view of Eq.~(\ref{Eq:ell}) this is the case if and only if $|\alpha| < \alpha_c(x_0)$ or $|\alpha-\pi| < \alpha_c(x_0)$, where
\begin{displaymath}
\alpha_c(x_0) := \arcsin\left( \frac{r(x_{th})}{r(x_0)} \right). 
\end{displaymath}
Notice that $0 < \alpha_c(x_0) \leq \pi/2$ and $\alpha_c(x_0) = \pi/2$ if and only if the initial point is at the throat, $x_0 = x_{th}$.
\item[(iii)] When $|\alpha| > \alpha_c(x_0)$ and $|\alpha-\pi| > \alpha_c(x_0)$, the 
maximum of the effective potential is larger than one and there is a turning point at $x_{turning}$ such that $r(x_{turning})^2 = \ell^2 = r(x_0)^2\sin^2\alpha > r(x_{th})^2$.
\item[(iv)] In the limiting case $\alpha = \alpha_c(x_0)$ or $\alpha = \alpha_c(x_0) + \pi$, the maximum of the effective potential is exactly one and $\ell = \pm r(x_{th})$. If $x_0 = x_{th}$, the solution describes an unstable circular orbit, for which the corresponding light ray winds around the throat. When $x_0\neq x_{th}$ the trajectories either converge to $x = \pm \infty$ or asymptotically approach the throat as $\lambda\to \infty$.
\end{enumerate}
From Eq.~(\ref{Eq:NullGeodesics}) and the definition $\ell = r(x)^2\dot{\varphi}$ one obtains the following expressions for the affine parameter $\lambda$ and the azimuthal angle $\varphi$:
\begin{displaymath}
\lambda(x_2) - \lambda(x_1) = \left| \int\limits_{x_1}^{x_2} 
 \frac{r(x) dx}{\sqrt{r(x)^2 - \ell^2}} \right|,\qquad
\varphi(x_2) - \varphi(x_1) = \ell\left| \int\limits_{x_1}^{x_2} 
 \frac{dx}{r(x)\sqrt{r(x)^2 - \ell^2}} \right|,
\end{displaymath}
for a path starting at $x_1$ and terminating at $x_2$, assuming that there is no turning point in between. From the first expression, it follows that the affine parameter diverges as $x_2\to \pm\infty$ in the cases (i)--(iii) above. In the limiting case (iv), where $\ell^2 = r(x_{th})^2$, the affine parameter also diverges when $x_2$ approaches the throat since near the throat $r(x) = r(x_{th}) + \frac{1}{2} r''(x_{th})(x-x_{th})^2 + {\cal O}(x-x_{th})^3$. Therefore, the effective potential is complete~\cite{ReedSimon80II} in the sense that each orbit exists for infinite values of the affine parameter. In particular, this implies that the wormhole spacetime~(\ref{Eq:WHMetric}) is null geodesic complete.

We may summarize the results so far in the following way: Let us introduce the effective phase space $\Gamma := \Real \times S^1$ of possible initial data $(x_0,e^{i\alpha})$ for the null geodesics. Consider the subset
\begin{displaymath}
\Gamma_\infty
 := \{ (x_0,e^{i\alpha} ) \in \Gamma : \lim\limits_{\lambda\to\infty} |x(\lambda)| = \infty \}
\end{displaymath}
of those data which give rise to a null geodesics which escapes to infinity in the future.  According to the analysis above, the complement of this set, describing future trapped null geodesics, is characterized by
\begin{displaymath}
\Gamma_{trapped} := \Gamma\setminus\Gamma_\infty
 = \left\{ (x_0,e^{i\alpha} ) \in \Gamma : 
 \alpha = \left\{ \begin{array}{rl}
 \alpha_c(x_0) & \hbox{for $x_0\leq x_{th}$} \\
 \alpha_c(x_0) + \pi & \hbox{for $x_0\geq x_{th}$}
 \end{array} \right\} \right\}.
\end{displaymath}
Since this set has measure zero in $\Gamma$, it follows that $\Gamma_{trapped}$ is \emph{unstable} in the sense that a generic perturbation of initial data $(x_0,\alpha)$ in this set leads to an orbit which reaches infinity in the future.

\begin{figure}[ht]
\centerline{\resizebox{10cm}{!}{\includegraphics{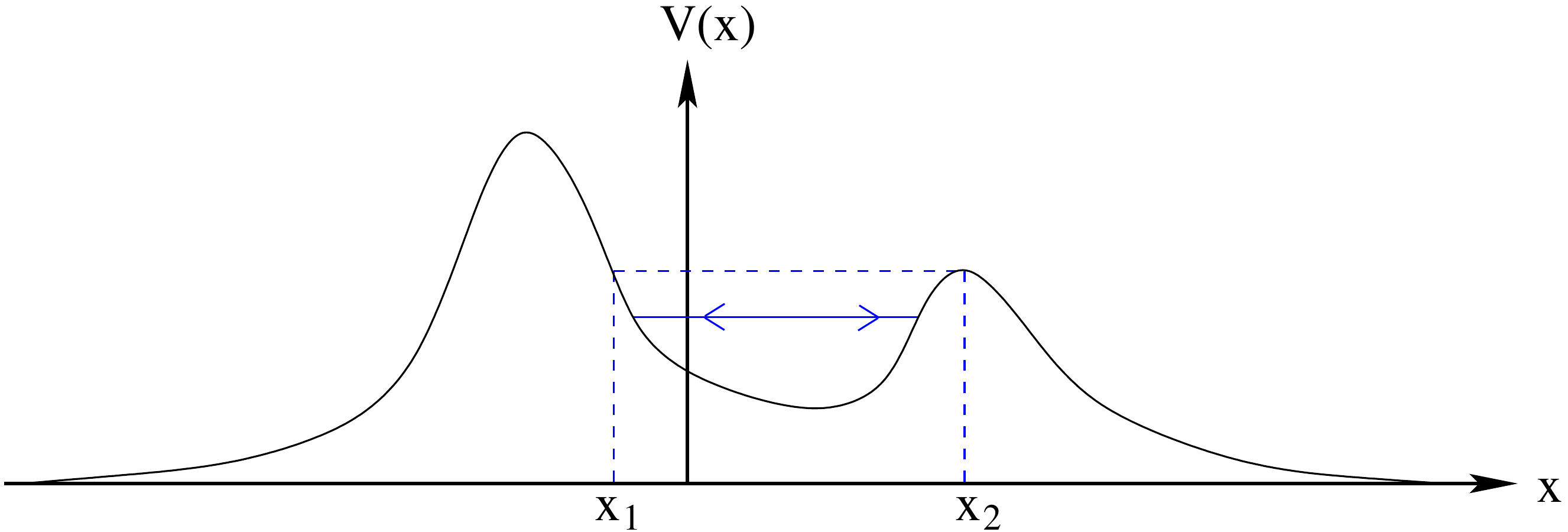}}}
\vspace{0.1cm}
\caption{An example for an effective potential with two local maxima and one local minimum, corresponding, respectively, to unstable and stable circular orbits. There is a potential well between $x_1$ and $x_2$ giving rise to oscillatory motion.}
\label{Fig:PotentialAsym}
\end{figure}

The results so far are based on the assumption that the areal function $r(x)$ is strictly convex, implying that the effective potential has a unique local maximum at the throat. However, consider now a background wormhole for which the effective potential exhibits the structure shown in Fig.~\ref{Fig:PotentialAsym}. Such potentials arise in wormhole models~\cite{nMtZ08} that were proposed as black hole foils~\cite{tDsS07}.
In contrast to the previous case where the effective potential has a unique local maximum, in the present case there is a potential well between $x_1$ and $x_2$ with a local minimum corresponding to stable circular orbits. Let us consider initial data $(x_0,e^{i\alpha})$ lying inside the well, that is $x_1 < x_0 < x_2$. It follows using similar arguments than in the previous case that for initial angles $\alpha$ satisfying $|\alpha| > \alpha_1(x_0)$ and $|\alpha-\pi| > \alpha_1(x_0)$ (ie. those angles sufficiently close to $\pi/2$ or $3\pi/2$) with
\begin{displaymath}
\alpha_1(x_0) = \arcsin\left( \frac{r(x_1)}{r(x_0)} \right),
\end{displaymath}
the motion is oscillatory, and the corresponding light rays follow a helicoidal type trajectory. Since these trajectories are trapped, the set
\begin{displaymath}
\left\{ (x_0,e^{i\alpha})\in \Gamma : x_1 < x_0 < x_2, 
 |\sin\alpha| > \frac{r(x_1)}{r(x_0)} \right\} \subset \Gamma_{trapped}
\end{displaymath}
is an open subset of $\Gamma_{trapped}$. As a consequence, $\Gamma_\infty = \Gamma\setminus\Gamma_{trapped}$ cannot be dense in $\Gamma$, and it is not true that a generic null geodesic escapes to infinity, like in the previous case.

\subsection{Timelike geodesics}

For timelike geodesics it is natural to choose the scaling freedom described below Eq.~(\ref{Eq:Geodesic}) such that $k=1$, implying that the affine parameter measures proper time. The initial four-velocity ${\bf u}$ can be parametrized in terms of an hyperbolic angle $\chi > 0$ and a conventional angle $\alpha$ according to
\begin{displaymath}
{\bf u} = \cosh\chi\frac{\partial}{\partial t} 
 + \sinh\chi\left( \cos\alpha\frac{\partial}{\partial x}
 + \frac{\sin\alpha}{r(x_0)}\frac{\partial}{\partial \varphi} \right),
\end{displaymath}
where $x_0$ is the $x$-coordinate of the initial point $p$. The conservation laws $\dot{t} = E$ and $r(x)^2\dot{\varphi} = \ell$ imply that
\begin{equation}
E = \cosh\chi,\qquad
\ell = r(x_0)\sin\alpha\sinh\chi,
\end{equation}
and consequently, the geodesic equation~(\ref{Eq:Geodesic}) can be rewritten as
\begin{equation}
\dot{x}^2 + \left( \frac{r(x_0)}{r(x)} \right)^2\sinh^2\chi\sin^2\alpha = \sinh^2\chi.
\label{Eq:TimeLikeGeodesics}
\end{equation}
Therefore, concerning the asymptotic limit of the timelike geodesics, we may draw similar conclusions than in the case of null geodesics. Now the effective phase space describing the set of possible initial data $(x_0,e^\chi,e^{i\alpha})$ is $\Gamma = \Real\times (1,\infty)\times S^1$, and we may write it as the disjoint union $\Gamma = \Gamma_{trapped} \,\dot{\cup}\, \Gamma_\infty$, were
\begin{displaymath}
\Gamma_\infty := \{ (x_0,e^\chi,e^{i\alpha} ) \in \Gamma
 : \lim\limits_{\lambda\to\infty} |x(\lambda)| = \infty \}
\end{displaymath}
and $\Gamma_{trapped} := \Gamma\setminus \Gamma_\infty$, as before. For an effective potential of the form displayed in Fig.~\ref{Fig:Potential} it follows again that $\Gamma_{trapped}$ is unstable, implying that there are no stable trapped timelike geodesics. In contrast to this, for the example shown in Fig.~\ref{Fig:PotentialAsym} the set $\Gamma_{trapped}$ contains the open subset
\begin{displaymath}
\left\{ (x_0,e^\chi,e^{i\alpha})\in \Gamma : x_1 < x_0 < x_2, \chi > 0, 
 |\sin\alpha| > \frac{r(x_1)}{r(x_0)} \right\},
\end{displaymath}
describing stable trapped timelike geodesics (planetary motion).

%%%%%%%%%%%%%%%%%%%%%%%%%%%%%%%%%%%%%%%%%%%%%
\section{Evolution of fields on a wormhole background}
\label{Sect:TestFields}
%%%%%%%%%%%%%%%%%%%%%%%%%%%%%%%%%%%%%%%%%%%%%

After having analyzed geodesic motion on the wormhole background~(\ref{Eq:WHMetric}), describing the evolution of photons or massive test particles, in this section we make a few remarks about the propagation of test fields on the background spacetime~(\ref{Eq:WHMetric}). For definiteness, we consider here a scalar field $\Phi$ of mass $\mu$ whose dynamics is described by the Klein-Gordon equation
\begin{equation}
-\nabla^\mu\nabla_\mu\Phi + \mu^2\Phi = 0.
\label{Eq:KG}
\end{equation}
When specialized to the spherically symmetric background~(\ref{Eq:WHMetric}), this can be reduced to a family of radial equations through a decomposition $\Phi = r(x)^{-1}\sum\limits_{\ell m} \phi_{\ell m}(t,x) Y^{\ell m}(\vartheta,\varphi)$ into spherical harmonics $Y^{\ell m}$,
\begin{equation}
\frac{\partial^2\phi_{\ell m}}{\partial t^2} - \frac{\partial^2\phi_{\ell m}}{\partial x^2}
 + V_{\ell\mu}(x)\phi_{\ell m} = 0,\qquad
V_{\ell\mu}(x) := \frac{r''(x)}{r(x)} + \frac{\ell(\ell+1)}{r(x)^2} + \mu^2.
\end{equation}
The first term, $r''(x)/r(x)$, in the effective potential $V_{\ell\mu}$ comes from the curvature of the spacetime background (notice that this term would be zero in the Minkowski case $r(x) = x$), the second term is the centrifugal term and is the quantum analog of the effective potential in Eq.~(\ref{Eq:Geodesic}) for the geodesic motion, while the last term is just inherited from the Klein-Gordon equation~(\ref{Eq:KG}).

For the Bronnikov-Ellis wormhole $r(x) = \sqrt{x^2 + b^2}$, for instance, the curvature term is $r''(x)/r(x) = b^2/r(x)^4$ and $V_{\ell\mu}$ is positive and has a global maximum at the throat $x=0$. Therefore, it describes a potential barrier and in principle, one can estimate transmission and reflection coefficients for scattering processes involving incoming radiation from one asymptotic flat end. It would be interesting to perform such an analysis and to compute the quasinormal modes which are expected to arise from the poles of the analytic continuation of these coefficients~\cite{hN99}.

Solutions of Eq.~(\ref{Eq:KG}) generated by smooth initial data with compact support remain bounded in time. This can be inferred from the existence of the conserved energy
\begin{equation}
E = \int\limits_{\Real\times S^2} \left[ \left( \frac{\partial\Phi}{\partial t} \right)^2
+ \gamma^{ij} D_i\Phi\cdot D_j\Phi + \mu^2\Phi^2 \right] \sqrt{\gamma} d^3 x,
\end{equation}
where $\gamma = dx^2 + r(x)^2 \left( d\vartheta^2 + \sin^2\vartheta\; d\varphi^2 \right)$ is the three-metric induced from ${\bf g}$ on the spatial, hypersurface-orthogonal slices $t=const$. This energy is the conserved quantity on such time slices, belonging to the conserved four-current $J^\mu = -T^\mu{}_\nu k^\nu$ which is constructed from the stress-energy tensor $T^\mu{}_\nu$ of the scalar field and the static Killing vector field $k = \partial/\partial t$ of the spacetime~(\ref{Eq:WHMetric}). Due to the orthonormality of the spherical harmonics, this energy reduces to a sum $E = \sum\limits_{\ell m} E_{\ell m}$ with
\begin{equation}
E_{\ell m} = \int\limits_{-\infty}^\infty \left[ 
 \left| \frac{\partial\phi_{\ell m}}{\partial t} \right|^2
 + r(x)^2\left|  \frac{\partial}{\partial x} \left( \frac{\phi_{\ell m}}{r(x)} \right) \right|^2
 + \frac{\ell(\ell+1)}{r(x)^2}|\phi_{\ell m}|^2 + \mu^2 |\phi_{\ell m}|^2 \right] dx,
\end{equation}
when decomposed into spherical harmonics. Since each term $E_{\ell m}$ is positive definite, one obtains a bound on the $L^2$ norm of $\phi_{\ell m}$ and their first order derivatives, and from this a uniform bound on $\phi_{\ell m}$.\footnote{This follows from a classical Sobolev estimate, see for instance~\cite{John-Book}. With a little bit more work one could probably also show that $\Phi$ itself is uniformly bounded on the spacetime~(\ref{Eq:WHMetric}) provided the initial data is smooth and has compact support.} 

In fact, similar conclusions can be obtained for Maxwell's equations on the wormhole background~(\ref{Eq:WHMetric}). It follows that the time evolution of scalar and electromagnetic fields from smooth initial data with compact support on such backgrounds stays bounded, and in this sense these fields are stable. However, as discussed in the next section, the situation may change drastically when the self-gravity of the fields is taken into account.

%%%%%%%%%%%%%%%%%%%%%%%%%%%%%%%%%%%%%%%%%%%%%
\section{Wormholes supported by scalar fields and their stability}
\label{Sect:Stability}
%%%%%%%%%%%%%%%%%%%%%%%%%%%%%%%%%%%%%%%%%%%%%

We consider here a family $\Phi = (\phi^A)$, $A=1,2,\ldots,m$, of  self-gravitating, noninteracting, minimally coupled, massless scalar fields. The action is
\begin{equation}
S[{\bf g},\Phi] = \int\left( -\frac{R}{16\pi G} 
 + \frac{1}{2} h_{AB}\nabla^\mu\phi^A\cdot\nabla_\mu\phi^B \right)\sqrt{-g}\, d^4 x,
\label{Eq:Action}
\end{equation}
where $R$ and $\nabla$ denote the Ricci scalar and the covariant derivate, respectively, associated to the spacetime metric ${\bf g}$, $G$ is Newton's constant and $(h_{AB}) = \mbox{diag}(-1,-1,\ldots,-1,+1,\ldots,+1)$ is a flat metric on internal space with $r$ negative and $s$ positive entries, corresponding to $r$ phantom and $s$ Klein-Gordon scalar fields. The corresponding field equations are
\begin{eqnarray}
R_{\mu\nu} &=& 8\pi G\, h_{AB}\nabla_\mu\phi^A \cdot \nabla_\nu\phi^B, 
\label{Eq:Einstein}\\
0 &=& \nabla^\mu\nabla_\mu\phi^A,
\label{Eq:KleinGordon}
\end{eqnarray}
where $R_{\mu\nu}$ is the Ricci tensor associated to ${\bf g}$. The action~(\ref{Eq:Action}) and field equations~(\ref{Eq:Einstein},\ref{Eq:KleinGordon}) are invariant with respect to the global transformations
\begin{equation}
\phi^A \mapsto \Lambda^A{}_B\phi^B + a^A,\qquad
\Lambda\in O(r,s),\quad {\bf a}\in \Real^m,
\label{Eq:Rotation}
\end{equation}
in the internal space. Using this symmetry, particular solutions of Eqs.~(\ref{Eq:Einstein},\ref{Eq:KleinGordon}) can be constructed by applying the transformation~(\ref{Eq:Rotation}) to a solution having only one component in internal space, $\Phi = (\phi,0,0,\ldots,0)$, in which case the system reduces to the Einstein-ghost scalar field equations whose solutions in the static, spherically symmetric sector are well known~\cite{hE73,kB73}. In fact, we claim that all the static, spherically symmetric wormhole solutions of Eqs.~(\ref{Eq:Einstein},\ref{Eq:KleinGordon}) can be obtained by this method. In order to show this, assume $({\bf g},\Phi)$ is such a solution. We can write the metric in the form
\begin{displaymath}
{\bf g} = -e^{-2a(x)} dt^2 + e^{2a(x)} dx^2 
 + r(x)^2 \left( d\vartheta^2 + \sin^2\vartheta\; d\varphi^2 \right),
\end{displaymath}
with a regular function $a(x)$ and $r(x)$ the areal function. In this parametrization, the wave equations~(\ref{Eq:KleinGordon}) reduce to
\begin{displaymath}
\frac{\partial}{\partial x}\left[ r(x)^2 e^{-2a(x)} \frac{\partial \phi^A}{\partial x} \right] = 0,
\qquad A=1,2,\ldots m,
\end{displaymath}
whose solutions are described by
\begin{displaymath}
\phi^A(x) = f(x)\alpha^A + \beta^A,\qquad
f(x) = \int\limits_{-\infty}^x\frac{e^{2a(y)}}{r(y)^2} dy,\qquad
A=1,2,\ldots m,
\end{displaymath}
with integration constants $\alpha^A$ and $\beta^A$. The important point to notice here, is that the function $f(x)$ is common to all the $\phi^A$'s. Therefore, we can apply the symmetry~(\ref{Eq:Rotation}) to transform $\phi^A$ to a vector in internal space of the form $\Phi' = (\phi',0,0,\ldots,0)$, provided we ensure that the vector $(\alpha^A)$ is timelike in internal space, $h_{AB}\alpha^A\alpha^B < 0$. In order to show that this must be the case, we consider the combination $R - 2R^x{}_x$ of the Einstein equations~(\ref{Eq:Einstein}) which yields
\begin{displaymath}
\frac{r'(x)^2}{r(x)^2} - 2a'(x)\frac{r'(x)}{r(x)} - \frac{e^{2a(x)}}{r(x)^2} 
 = 4\pi G\, h_{AB}\left( \frac{\partial\phi^A}{\partial x} \right)
 \left( \frac{\partial\phi^B}{\partial x} \right).
\end{displaymath}
Evaluating at the throat $x=x_{th}$ and using $r'(x_{th})=0$ combined with the regularity of $a(x)$, it follows that $4\pi G\, h_{AB}\alpha^A\alpha^B f'(x_{th})^2 = -e^{2a(x_{th})}/r(x_{th})^2$, which shows that $f'(x_{th})\neq 0$ and that $h_{AB}\alpha^A\alpha^B$ is negative. Therefore, as a consequence of the symmetry~(\ref{Eq:Rotation}), any static, spherically symmetric wormhole solution of Eqs.~(\ref{Eq:Einstein},\ref{Eq:KleinGordon}) is equivalent to a static, spherically symmetric solution of the Einstein-ghost scalar field theory. Since the latter are known to be linearly unstable~\cite{jGfGoS09a} it follows that any static, spherically symmetric wormhole solution of the theory involving $r$ ghost and $s$ Klein-Gorden scalar fields is unstable with respect to linear fluctuations of the metric and scalar fields. The numerical results for the nonlinear evolution in~\cite{hSsH02,jGfGoS09b,aDnKdNiN08} show that such wormholes may collapse to a Schwarzschild black hole.

In conclusion, this model shows that the inclusion of one or more Klein-Gordon fields to the system does not lead to wormholes which are linearly stable. This result, in combination with the effect of the electric charge on the stability \cite{jGfGoS09c}, suggests that it might be unlikely to stabilize a wormhole by adding ordinary matter.

%%%%%%%%%%%%%%%%%%%%%%%%%%%%%%%%%%%%%%%%%%%%%
\section{Conclusions}
\label{Sect:Conclusions}
%%%%%%%%%%%%%%%%%%%%%%%%%%%%%%%%%%%%%%%%%%%%%

In this work we have presented some basic properties of static, spherically symmetric wormhole spacetimes. For the wormhole models given in Eq.~(\ref{Eq:WHMetric}) we have shown null and timelike geodesic completeness. Furthermore, we have shown that depending upon the structure of the effective potential, there might exist trapped timelike and null geodesics winding around the throat. In contrast to the Schwarzschild black hole, cases are exhibited involving stable trapped null geodesics. As far as fields are concerned, we have shown that a test scalar field on the wormhole background~(\ref{Eq:WHMetric}) remains bounded in time for smooth initial data with compact support. Finally, we consider a mixture of self-gravitating ghost and Klein-Gordon scalar fields and prove that the resulting static, spherically symmetric wormholes are linearly unstable. In this context, it is interesting to mention that the nonlinear stability of Minkowski spacetime in the Einstein-scalar field model has been established not only for a Klein-Gordon field but also for a ghost scalar field (see the comments and references in appendix B.5 of~\cite{mDiR08}). Therefore, it seems that both the \emph{self-gravity} and the \emph{topology} are key factors leading to the instability of the wormholes.

%%%%%%%%%%%%%%%%%%%%%%%%%%%%%%%%%%%%%%%%%%%%%
\begin{theacknowledgments}
%%%%%%%%%%%%%%%%%%%%%%%%%%%%%%%%%%%%%%%%%%%%%

We thank E. Chaverra, N. Montelongo-Garc\'{\i}a and N. Ortiz for discussions. OS also wishes to thank H. Friedrich for stimulating discussions. This work was supported in part by CIC Grants No. 4.7 and No. 4.19 to Universidad Michoacana.
\end{theacknowledgments}

\bibliographystyle{aipproc}  
\bibliography{refs}

\begin{thebibliography}{31}
\expandafter\ifx\csname natexlab\endcsname\relax\def\natexlab#1{#1}\fi
\providecommand{\enquote}[1]{``#1''}
\expandafter\ifx\csname url\endcsname\relax
  \def\url#1{\texttt{#1}}\fi
\expandafter\ifx\csname urlprefix\endcsname\relax\def\urlprefix{URL }\fi
\providecommand{\eprint}[2][]{\url{#2}}

\bibitem[Morris and Thorne(1988)]{mMkT88}
M.~Morris, and K.~Thorne, \emph{Am. J. Phys.} \textbf{56}, 395--412 (1988).

\bibitem[Morris et~al.(1988)]{mMkTyU88}
M.~Morris, K.~Thorne, and U.~Yurtsever, \emph{Phys. Rev. Lett.} \textbf{61},
  1446--1449 (1988).

\bibitem[Visser(1995)]{Visser-Book}
M.~Visser, \emph{Lorentzian wormholes. {F}rom {E}instein to {H}awking},
  American Institute of Physics, Woodbury, New York, 1995.

\bibitem[Friedman et~al.(1993)]{jFkSdW93}
J.~Friedman, K.~Schleich, and D.~Witt, \emph{Phys. Rev. Lett.} \textbf{71},
  1486--1489 (1993).

\bibitem[Epstein et~al.(1965)]{hEvGaJ65}
H.~Epstein, V.~Glaser, and A.~Jaffe, \emph{Nuovo Cim.} \textbf{36}, 1016--1022
  (1965).

\bibitem[Ford(1978)]{lF78}
L.~Ford, \emph{Proc. Roy. Soc. Lond. A} \textbf{364}, 227--236 (1978).

\bibitem[Ford(1991)]{lF91}
L.~Ford, \emph{Phys. Rev. D} \textbf{43}, 3972--3978 (1991).

\bibitem[Ford and Roman(1996)]{lFtR96}
L.~Ford, and T.~A. Roman, \emph{Phys. Rev. D} \textbf{53}, 5496--5507 (1996).

\bibitem[Yurtsever(1995)]{uY05}
U.~Yurtsever, \emph{Phys.Rev. D} \textbf{52}, 564--568 (1995).

\bibitem[Flanagan and Wald(1996)]{eFrW96}
E.~Flanagan, and R.~Wald, \emph{Phys. Rev. D} \textbf{54}, 6233--6283 (1996).

\bibitem[Fewster and Roman(2005)]{cFtR05}
C.~Fewster, and T.~Roman, \emph{Phys.Rev. D} \textbf{72}, 044023 (2005).

\bibitem[Kanti et~al.(2011)]{pKbKjK11}
P.~Kanti, B.~Kleihaus, and J.~Kunz, \emph{Phys.Rev.Lett.} \textbf{107}, 271101
  (2011).

\bibitem[Kanti et~al.(2012)]{pKbKjK12}
P.~Kanti, B.~Kleihaus, and J.~Kunz, \emph{Phys. Rev. D} \textbf{85}, 044007
  (2012).

\bibitem[Montelongo~Garc\'{\i}a and Lobo(2011)]{nMfL11}
N.~Montelongo~Garc\'{\i}a, and F.~Lobo, \emph{Class. Quantum Grav.}
  \textbf{28}, 085018 (2011).

\bibitem[Gonz\'alez et~al.(2009{\natexlab{a}})]{jGfGoS09a}
J.~A. Gonz\'alez, F.~S. Guzm\'an, and O.~Sarbach, \emph{Class. Quantum Grav.}
  \textbf{26}, 015010 (2009{\natexlab{a}}).

\bibitem[Shinkai and Hayward(2002)]{hSsH02}
H.~Shinkai, and S.~Hayward, \emph{Phys. Rev. D} \textbf{66}, 044005 (2002).

\bibitem[Gonz\'alez et~al.(2009{\natexlab{b}})]{jGfGoS09b}
J.~A. Gonz\'alez, F.~S. Guzm\'an, and O.~Sarbach, \emph{Class. Quantum Grav.}
  \textbf{26}, 015011 (2009{\natexlab{b}}).

\bibitem[Doroshkevitch et~al.(2008)]{aDnKdNiN08}
A.~Doroshkevitch, N.~Kardashev, D.~Novikov, and I.~Novikov, \emph{Astron. Rep.}
  \textbf{52}, 616--622 (2008).

\bibitem[Gonz\'alez et~al.(2009{\natexlab{c}})]{jGfGoS09c}
J.~A. Gonz\'alez, F.~S. Guzm\'an, and O.~Sarbach, \emph{Phys. Rev. D}
  \textbf{80}, 024023 (2009{\natexlab{c}}).

\bibitem[Shatskii et~al.(2008)]{aSiNnK08}
A.~Shatskii, I.~Novikov, and N.~Kardashev, \emph{Physics -- Uspekhi}
  \textbf{51}, 457--464 (2008).

\bibitem[Novikov et~al.(2009)]{dNaDiNaS09}
D.~Novikov, A.~Doroshkevich, I.~Novikov, and A.~Shatskii, \emph{Astron. Rep.}
  \textbf{53}, 1079--1085 (2009).

\bibitem[Sarbach and Zannias(2010)]{oStZ10}
O.~Sarbach, and T.~Zannias, \emph{Phys. Rev. D} \textbf{81}, 047502 (2010).

\bibitem[Novikov and Shatskiy(2012)]{iNaS12}
I.~Novikov, and A.~Shatskiy  (2012), \eprint{arXiv:1201.4112}.

\bibitem[Ellis(1973)]{hE73}
H.~Ellis, \emph{J. Math. Phys.} \textbf{14}, 104--118 (1973).

\bibitem[Bronnikov(1973)]{kB73}
K.~Bronnikov, \emph{Acta Phys. Polonica B} \textbf{4}, 251--266 (1973).

\bibitem[Reed and Simon(1980)]{ReedSimon80II}
M.~Reed, and B.~Simon, \emph{Methods of Modern Mathematical Physics, Vol. II:
  Fourier Analysis, Self-Adjointness}, Academic Press, San Diego, 1980.

\bibitem[Garc\'{\i}a and Zannias(2008)]{nMtZ08}
N.~M. Garc\'{\i}a, and T.~Zannias, \emph{Phys. Rev. D} \textbf{78}, 064003
  (2008).

\bibitem[Damour and Solodukhin(2007)]{tDsS07}
T.~Damour, and S.~Solodukhin, \emph{Phys. Rev. D} \textbf{76}, 024016 (2007).

\bibitem[Nollert(1999)]{hN99}
H.-P. Nollert, \emph{Class. Quantum Grav.} \textbf{16}, R159--R216 (1999).

\bibitem[John(1982)]{John-Book}
F.~John, \emph{Partial differential equations}, Springer-Verlag, 1982.

\bibitem[Dafermos and Rodnianski(2008)]{mDiR08}
M.~Dafermos, and I.~Rodnianski  (2008), \eprint{arXiv:0811.0354}.

\end{thebibliography}
%%%%%%%%%%%%%%%%%%%%%%%%%%%%%%%%%%%%%%%%%%%%%

\end{document}